\documentclass[prb,preprint,aps,longbibliography,groupedaddress]{revtex4-1}

\usepackage{bbm}
\usepackage{graphicx}
\usepackage{subfigure}
\usepackage{amsmath}
\usepackage{booktabs}
\usepackage{ulem}

\begin{document}

\title{Balanced ternary formalism of second quantization}

\author{Yao Yao\footnote{\url{yaoyao2016@scut.edu.cn}}}

\address{
Department of Physics and State Key Laboratory of Luminescent Materials and Devices, South China University of Technology, Guangzhou 510640, China}

\date{\today}

\begin{abstract}
We construct a second-quantized representation with a structure of balanced ternary formalism, which involves three substances in organic molecular materials, namely electron, hole and charge-transfer exciton, into a uniform framework. The quantum thermodynamic of excitons is investigated in a closed and compact manner, benefitting from the interplay of the three substances. In order to be friendly with quantum simulations, the interactions among them are all described with unitary transformations. Significantly, the nonconserving dynamics of particle numbers, such as the generation of charge current and the exciton fission in organic semiconductors, is consistently expressed by this unitary formalism on the basis of bosonic coherent states. The spin degree of freedom is further taken into account, and an exotic molecular ferromagnetic ordering is induced in a specific configuration of excitons. This balanced ternary formalism establishes a solid bridge to connect thermodynamics and quantum simulations.
\end{abstract}

\maketitle

\section{Introduction}

Conventional second quantization for fermions relies on binary state spaces, with each binary system denoting their occupied and unoccupied states, respectively. In this formalism, there are two ways to describe electron and hole. The first is to set occupied state to be electron and unoccupied one for hole. This implies the total number of electron and hole will never be changed, so it can not be applied to describe the nonconserving systems. The second way is to assign two individual binary systems, one of which is for electron and the other for hole. By this way, the electron and hole are not mutual exclusive and the coexistence of them at a single spatial place is commonly allowed, which is not the realistic case, as for example a hydrogen can normally have three valence states $+1,0,-1$, and the $\pm1$ states can never simultaneously appear in a single atom. In addition, the electron and hole do not share the same ground state by the second construction. The recombination of them can not be nicely expressed, since the associative creation and annihilation of photogenerated charges are ill defined without a compact commutation relation. In this context, a balanced ternary formalism taking electron, hole and exciton into a uniform framework is necessarily demanded.

There are a number of realistic materials that are naturally disordered due to their structural diversity \cite{disorder}. Organic molecular aggregates, polymers, amorphous crystals, glasses and so on constitute this huge family of disordered materials. The structural disorders are normally fixed during synthesis and can hardly be changed temporally, so they can be regarded as static disorders, leading to localization of electronic orbitals. It is then not available to write down a translational invariant Hamiltonian for them with its eigenstate being plane wave. The quantum coherence between these localized orbitals, on the other hand, is generated and eliminated temporally by dynamic disorders stemming from molecular vibrations (phonons) \cite{DynDis}. In a common treatment, the system will be exerted to an open environment and the dynamics is described by some non-unitary evolution \cite{master}. In the materials with heavy elements or rigid backbones, however, the influence of dynamic disorders should be largely weakened \cite{rigid}. Subsequently, the coherent length, i.e. the distance that two localized orbitals hold their coherence in between, could be much longer than the localization length. For example, in some rigid molecular aggregates the electrons are normally localized on each molecules, but the coherent length could be on the order of ten molecules \cite{ten1,ten2}.

This mismatch of localization and nonlocal coherence can not be consistently described by binary systems. Let us consider a localized system with the local integrals of motion $\sigma_i$. If they are binary systems, one can solely generate the nonlocal coherence by constructing Ising-like interaction terms as $\sigma_i\sigma_j$ \cite{local}, which is however in a classical fashion. In a quantum Ising model, the localization must be broken by some kind of exchange, leading to the incompatibility of the localization and nonlocal coherence. Quantum ternary systems (qutrits) therefore serve as the minimal model to be compatible with them. One can construct various interaction terms with respect to different states of each qutrit and let them compete with each other. In the negative-temperature setup, e.g., the qutrit is the minimal system to concurrently contact with two thermal baths via different two states of it and generate the synthetic negative temperatures \cite{negative,negative2}. Consequently, the ternary formalism has got great advantage in describing the quantum thermodynamics.

Excitons are important composite particles in condensed matter, which attract much attention in recent years \cite{ExRev}. In particular, the condensation of excitons acts as an appealing realization of Bose-Einstein condensation \cite{conden1,conden2}. Commonly speaking, excitons can be thought as the tightly-binding electron-hole pairs, which represent the nonlocal coherence between two localized charge carriers, so they are not easy to be explicitly described by the binary formalism. In the past, people normally treat the excitons as individual excited states and empirically adopt projectors onto these states to express the generation, recombination, dissociation and transition of excitons. By this treatment, the quantum dynamics and thermodynamics of excitons usually appear as trivial composition of Rabi oscillations between ground and excited states. In addition, some valley excitons are naturally non-Hermitian systems exerting to external photons \cite{valley}, so that the open dynamics is intuitively adopted as well. In order to comprehend the quantum and statistical features of excitons in a closed and compact fashion, therefore, the ternary system also turns out to be the minimal model.

Ternary systems are friendly to the quantum computations and simulations \cite{ternary}. Benefitting from the higher information density, the qutrits enable logarithmical reduction of the unit number than the qubit. In a technical manner, the quantum control usually employs photons of light or microwave whose angular momentum is more matched with qutrit. More importantly, quantum units are commonly quantum constrained systems with great localization, and quantum logic gates are unitary transformations that correlate these local systems. So quantum computations just utilize the quantum coherence between localized quantum units as the resource. Subsequently, there have been a number of researches on the realization of quantum ternary logic \cite{TerExp}. In this context, we propose a balanced ternary formalism which is more compatible with the conventional second quantization representation and could be straightforwardly extended to the quantum simulations.

The paper is organized as follows. In the next Section we give the basic definition of operators on a single qutrit. Section III is devoted to describe the basic formalism of excitons, as well as their thermodynamic features. The interactions between excitons and charge carriers are discussed in Section IV, where the exciton fission is studied as well. The spin-related issues are placed in Section V, and the exciton-induced ferromagnetic orderings are comprehended. Section VI presents a brief discussion on the Hamiltonian. The concluding remarks are addressed in the last Section.

\section{Formalism of single qutrit}

We first define two sets of creation (annihilation) fermionic operators $\hat{c}^{\dag}~(\hat{c})$ and $\hat{d}^{\dag}~(\hat{d})$. Both of them satisfy canonical anti-commutation relation, and we can simply regard the former set as for spinless electrons and the latter for holes. In realistic materials, the electron can be considered as the carrier on the conduction band minimum (CBM) or the lowest unoccupied molecular orbital (LUMO), and the hole is on the valance band maximum (VBM) or the highest occupied molecular orbital (HOMO). Acting these operators onto the usual closed-shell ground state $|0\rangle$ results in four bases, but here we are mainly focusing on three of them. Concretely, we construct balanced ternary states as
\begin{equation}
|+\rangle=\hat{c}^{\dag}|0\rangle,~~|\circ\rangle=\frac{1}{\sqrt{2}}(|0\rangle+\hat{c}^{\dag}\hat{d}|0\rangle),~~|-\rangle=\hat{d}|0\rangle.
\end{equation}
We do not take the state $|\circ'\rangle=\frac{1}{\sqrt{2}}(|0\rangle-\hat{c}^{\dag}\hat{d}|0\rangle)$ into the framework, because in the following derivations it can be safely screened by the proper arrangement of operators. From here on, we call $|\pm\rangle$ the electron or hole state and $|\circ\rangle$ the vacuum state. In our theory, when both electron and hole reside exactly on the same place, they will ultrafast recombine. Namely, in a thermal environment without continuous photonic pumping, on-site excitons could not stably exist, so $\hat{c}^{\dag}\hat{d}|0\rangle$ is equivalent to the vacuum. Throughout this paper, therefore, we solely discuss the charge-transfer (CT) excitons \cite{CT}.

The key input of the present formalism is to introduce two new sets of operators $\hat{f}^{\dag}~(\hat{f})$ and $\hat{g}^{\dag}~(\hat{g})$, which also satisfy the anti-commutation relation. For this aim, the definition writes
\begin{eqnarray}
&&\hat{f}^{\dag}=\frac{1}{\sqrt{2}}(\hat{c}^{\dag}+\hat{d}^{\dag}),~~\hat{f}=\frac{1}{\sqrt{2}}(\hat{c}+\hat{d}),\nonumber\\
&&\hat{g}^{\dag}=\frac{1}{\sqrt{2}}(\hat{c}^{\dag}-\hat{d}^{\dag}),~~\hat{g}=\frac{1}{\sqrt{2}}(\hat{c}-\hat{d}).
\end{eqnarray}
It is easy to derive that
\begin{eqnarray}
&&\hat{f}^{\dag}|-\rangle=|\circ\rangle,~~\hat{f}|\circ\rangle=|-\rangle,~~\hat{f}|+\rangle=|\circ'\rangle,\nonumber\\
&&\hat{g}^{\dag}|\circ\rangle=|+\rangle,~~\hat{g}^{\dag}|-\rangle=-|\circ'\rangle,~~\hat{g}|+\rangle=|\circ\rangle.
\end{eqnarray}
and
\begin{eqnarray}
&&\hat{f}^{\dag}\hat{f}|+\rangle=|+\rangle,~~\hat{f}^{\dag}\hat{f}|\circ\rangle=|\circ\rangle,~~\hat{f}\hat{f}^{\dag}|-\rangle=|-\rangle,\nonumber\\
&&\hat{g}^{\dag}\hat{g}|+\rangle=|+\rangle,~~\hat{g}\hat{g}^{\dag}|\circ\rangle=|\circ\rangle,~~\hat{g}\hat{g}^{\dag}|-\rangle=|-\rangle.
\end{eqnarray}
Actions of these operators on other states not listed here equal to zero. One can find that, only $\hat{g}^{\dag}$ and $\hat{f}$ can produce electron and hole from vacuum state, so we can call them ``real" operators. While $\hat{f}^{\dag}$ and $\hat{g}$ can merely assist to quench existed hole and electron, so they can be regarded as ``image" operators from an imaginary thermal environment entangled with the real system. As a result, these two sets of operators fulfill anti-commutation relation, namely,
\begin{eqnarray}
\{\hat{f},\hat{f}^{\dag}\}=\{\hat{g},\hat{g}^{\dag}\}=1,
\end{eqnarray}
and others are anti-commutative. Obviously, $\hat{f}^2=\hat{f}^{\dag2}=\hat{g}^2=\hat{g}^{\dag2}=0$ as usual. Although satisfying anti-commutation relation, we can not call them fermionic operators, as they act on a three-state system (qutrit). They are also different from the raising and lowering operators for the triplet state of spin one, as that in the AKLT model \cite{AKLT}, since the vacuum state in our formalism is not triplet.

In essence, $\hat{f}^{\dag}$ and $\hat{g}^{\dag}$ mean to create an electron or annihilate a hole, so that the whole number of particle is increased by one in each action, considering the number of hole is denoted by negative integers. And $\hat{f}$ and $\hat{g}$ are for the reverse case. We can thus define a number operator
\begin{eqnarray}
\hat{n}=\hat{f}^{\dag}\hat{f}-\hat{g}\hat{g}^{\dag}.
\end{eqnarray}
For $|+\rangle,|\circ\rangle,|-\rangle$, the eigenvalues are $1,0,-1$, respectively, so it is safe to regard these balanced ternary numbers as the particle numbers of electron, vacuum and hole. The commutation relations then write
\begin{eqnarray}
&&[\hat{n},\hat{f}^{\dag}]=\hat{f}^{\dag},~~[\hat{n},\hat{f}]=-\hat{f},\nonumber\\
&&[\hat{n},\hat{g}^{\dag}]=\hat{g}^{\dag},~~[\hat{n},\hat{g}]=-\hat{g},
\end{eqnarray}
equivalent to the normal cases of fermions. It is thus convenient to construct a second-quantized representation with the same form as the canonical ones.

One would be noticing that, most of the dual actions are vanishing, but $\hat{f}\hat{g}^{\dag}|\circ\rangle=-\hat{g}^{\dag}\hat{f}|\circ\rangle=|\circ'\rangle$, out of the bases. So we have to avoid these actions, that is, two real particles (with the same spin) can not coexist at exactly a same site, also known as the Pauli exclusion principle.

Another interesting actions emerge as
\begin{eqnarray}
\hat{g}^{\dag}\hat{f}^{\dag}|-\rangle=|+\rangle,~~\hat{f}\hat{g}|+\rangle=|-\rangle.
\end{eqnarray}
This reveals the exotic physical meaning of two individual sets of operators acting on a single qutrit. To this end, if we define a so-called thermally excited state as
\begin{eqnarray}
|\alpha\rangle=\frac{1}{\sqrt{2\cosh \alpha}}(e^{\alpha/2}|-\rangle+e^{-\alpha/2}|+\rangle),
\end{eqnarray}
where following the usual definition, $\alpha~(=-\mu/k_{\rm B}T)$ is a dimensionless parameter that can be thought as chemical potential difference between electron and hole. For vacuum we set its chemical potential, namely the ground-state energy, to be much smaller than that of electron and hole, so it does not have component in this excited state. Then it can be derived that
\begin{eqnarray}
\langle\alpha|\hat{f}^{\dag}\hat{f}|\alpha\rangle=\langle\alpha|\hat{g}^{\dag}\hat{g}|\alpha\rangle=\frac{1}{e^{2\alpha}+1},
\end{eqnarray}
equivalent to the Fermi-Dirac distribution. We further define $\sin\theta=(e^{2\alpha}+1)^{-\frac{1}{2}}$, so $|\alpha\rangle$ is nothing but
\begin{eqnarray}\label{bog}
|\theta\rangle=\cos\theta|-\rangle+\sin\theta|+\rangle=\exp(i\theta\hat{\gamma})|-\rangle,
\end{eqnarray}
with $\hat{\gamma}=i(\hat{f}\hat{g}-\hat{g}^{\dag}\hat{f}^{\dag})$. That is, from $\theta=\pi/4$, we can conduct a unitary Bogoliubov transformation with respect to $\hat{\gamma}$ to inject net electron or hole with the state always being pure. According to the thermofield dynamics \cite{thermo1,thermo2}, the operators $\hat{g}^{\dag}$ and $\hat{f}^{\dag}$ can then be regarded as dually entangled system and environment operators. Consequently, a three-state system is irreducible for describing quantum thermodynamics.

\section{Bosonization on a chain of qutrits}

By representing electron and hole with individual bases, it is the most convenience to construct the formalism of excitons by bosonization. We can define a set of bosonic operators for excitons, instead of merely using the projectors as usual. We can consistently discuss the thermodynamics of excitons which has not attracted sufficient attention before.

\subsection{Formalism of charge-transfer excitons}

We now construct a one-dimensional open chain consisting of $L$ sites, and each site is a qutrit with states $|+\rangle,|\circ\rangle,|-\rangle$. It follows the usual rule of second quantization. For example,
\begin{eqnarray}
&&\hat{g}^{\dag}_{\mu}|\cdot\cdot\cdot\circ_{\mu}\cdot\cdot\cdot\rangle=\pm|\cdot\cdot\cdot+_{\mu}\cdot\cdot\cdot\rangle,\nonumber\\
&&\hat{f}_{\mu}|\cdot\cdot\cdot\circ_{\mu}\cdot\cdot\cdot\rangle=\pm|\cdot\cdot\cdot-_{\mu}\cdot\cdot\cdot\rangle,
\end{eqnarray}
where dots denote other sites, and even occupied states of electron or hole in front of $\mu$-th site give positive sign and odd occupation for negative sign, respectively. In principle, the following derivations can be directly extended to higher dimensions. For a realistic case, one can think the chain as a molecular aggregate, a polymer or a disordered crystal, and each site representing a molecule, a monomer or a localization region on them offers active orbitals that can be occupied by either an electron or a hole.

Motivated by the fact that an electron and a hole form a CT exciton, which is a boson, we construct a set of bosonic operators as
\begin{eqnarray}
\hat{a}^{\dag}_{\mu,\nu}=\hat{g}^{\dag}_{\mu}\hat{f}^{\dag}_{\mu}\hat{f}_{\mu}\hat{\xi}\hat{f}_{\nu}\hat{g}_{\nu}\hat{g}^{\dag}_{\nu},~
\hat{a}_{\mu,\nu}=\hat{f}^{\dag}_{\nu}\hat{g}_{\nu}\hat{g}^{\dag}_{\nu}\hat{\xi}\hat{g}_{\mu}\hat{f}^{\dag}_{\mu}\hat{f}_{\mu},
\end{eqnarray}
where $\mu$ and $\nu$ are the site index, and $\hat{\xi}=\prod_{\delta=\nu+1}^{\mu-1}\exp(i\pi\hat{n}_{\delta})$ is the string operator between them to eliminate the negative sign during swapping. Herein, we have to introduce additional number operators $\hat{f}^{\dag}_{\mu}\hat{f}_{\mu}$ and $\hat{g}_{\nu}\hat{g}^{\dag}_{\nu}$ especially into the creation operators to eliminate unwanted actions to $|\circ'\rangle$ (the annihilations do not need), so the generation of excitons is necessary to be assisted by image operators. Considering only $\hat{g}^{\dag}$ and $\hat{f}$ have nonvanishing actions on vacuum, this construction is unique. It is worth noting that, this set of operators can be mapped to the nonlocal projectors of Rydberg atoms, which can be well simulated by state-of-the-art quantum platforms \cite{hop}.

By this construction, the nonvanishing actions can then be written as
\begin{eqnarray}
&&\hat{a}^{\dag}_{\mu,\nu}|\cdot\cdot\cdot\circ_{\mu}\cdot\cdot\cdot\circ_{\nu}\cdot\cdot\cdot\rangle=|\cdot\cdot\cdot+_{\mu}\cdot\cdot\cdot-_{\nu}\cdot\cdot\cdot\rangle,\nonumber\\
&&\hat{a}_{\mu,\nu}|\cdot\cdot\cdot+_{\mu}\cdot\cdot\cdot-_{\nu}\cdot\cdot\cdot\rangle=|\cdot\cdot\cdot\circ_{\mu}\cdot\cdot\cdot\circ_{\nu}\cdot\cdot\cdot\rangle,
\end{eqnarray}
with dots denoting that other sites can be at any states. Given a vacuum state on the chain $|\circ\rangle$, namely all sites are vacuum, then we have
\begin{eqnarray}
(\hat{a}^{\dag}_{\mu,\nu}\hat{a}_{\mu,\nu})\hat{a}^{\dag}_{\mu,\nu}|\circ\rangle=\hat{a}^{\dag}_{\mu,\nu}|\circ\rangle.
\end{eqnarray}
Here for simplicity we use the same symbol of vacuum state with that of a single site which does not matter. This result means $\hat{m}_{\mu,\nu}=\hat{a}^{\dag}_{\mu,\nu}\hat{a}_{\mu,\nu}$ can be regarded as a number operator of these CT excitons, and $\hat{a}^{\dag}_{\mu,\nu}|\circ\rangle$ is the eigenstate with the particle number being one. It is easy to check that,
\begin{eqnarray}\label{commu}
[\hat{a}^{\dag}_{\mu,\nu},\hat{a}^{\dag}_{\mu',\nu'}]=[\hat{a}_{\mu,\nu},\hat{a}_{\mu',\nu'}]=0,~~\hat{a}^{\dag2}_{\mu,\nu}|\circ\rangle=0
\end{eqnarray}
for any sites with $\mu\neq\mu'$ or $\nu\neq\nu'$, so the excitons are hard-core bosons. These bosonic operators also do not change the total number of electron and hole, that is, $[\hat{a}^{\dag}_{\mu,\nu},\hat{n}]=[\hat{a}_{\mu,\nu},\hat{n}]=0$ with $\hat{n}$ being the total number of fermions, namely $\hat{n}=\sum_{\mu}\hat{n}_{\mu}$.

One may be doubting that, although it possesses the bosonic feature, this exciton seems not a tightly-binding electron-hole pair as usual picture, since one can just simply construct a hopping operator $\hat{h}_{\mu',\mu}=\hat{g}^{\dag}_{\mu'}\hat{f}^{\dag}_{\mu'}\hat{f}_{\mu'}\hat{f}_{\mu}\hat{g}_{\mu}\hat{g}^{\dag}_{\mu}$ to freely move the electron in the exciton from $\mu$ to $\mu'$. We thus have to further consider the quantum coherence or entanglement between electron and hole to make sure the exciton is an individual substance. To this end, we define both $|+-\rangle$ and $|-+\rangle$ as the two bases of the same CT exciton to preserve the parity, and a current operator ($\hbar=1$ throughout this paper)
\begin{eqnarray}\label{current}
\hat{J}_{\mu,\nu}=-i(\hat{f}_{\mu}\hat{g}_{\mu}\hat{g}^{\dag}_{\nu}\hat{f}^{\dag}_{\nu}-{\rm h.c.})
\end{eqnarray}
can make transition between them. By acting a unitary transformation $e^{i\Theta\hat{J}}$ on $|+-\rangle$, it will be changed to the entangled state, namely
\begin{eqnarray}
e^{i\Theta\hat{J}}|+-\rangle=\cos\Theta|+-\rangle+\sin\Theta|-+\rangle.
\end{eqnarray}
The phase factor $\Theta$ can be thought as the polarization direction of exciton, and if there is an external electric field, it will be polarized to a certain direction, very similar with the case of spin half in magnetic field. Throughout this work, we do not consider the non-Hermitian effect of excitons which may be induced by for example valley polarization \cite{valley}. Due to the entanglement between electron and hole, the recombination, dissociation and transfer of the excitons demand for external energy input. For instance, the hopping operator $\hat{h}$ can no longer freely move the electron without breaking the coherence.

\subsection{Formalism of $y$-excitons}

For any $\hat{a}_{\mu',\nu'}\hat{a}^{\dag}_{\mu,\nu}$ with $\mu-\nu\neq\mu'-\nu'$, the action on vacuum state $|\circ\rangle$ vanishes, and actions on other states will be discussed below. It means, in terms of vacuum, merely the distance $y=|\mu-\nu|$ matters. We then call all CT excitons with $y$ distance as $y$-excitons, and $y$ is the coherent length of the long-range CT states, strongly depending on the thermal environment. The long-range CT excitons serve as the primary source of photo-generated charges \cite{CT}. In a normal molecular aggregate, especially with radicals, the coherent length is of the order around ten molecules \cite{ten1}, which is sufficient to justify the following discussions. Rather, the longer the electron and hole, the faster the decoherence between them. $L-1$ is assumed to be the longest coherence length for stable $y$-excitons.

In order to generate more $y$-excitons than one, we can sum up all operators on possible sites. We define the creation and annihilation for $y$-exciton as
\begin{eqnarray}
&&\hat{b}^{\dag}_y=\zeta_y\sum_{\mu-\nu=y}e^{i\Theta\hat{J}_{\mu,\nu}}\hat{a}^{\dag}_{\mu,\nu},\nonumber\\
&&\hat{b}_y=\zeta_y\sum_{\mu-\nu=y}\hat{a}_{\mu,\nu}e^{-i\Theta\hat{J}_{\mu,\nu}},
\end{eqnarray}
where $\zeta_y=(L-y-2(m_y-1))^{-1/2}$ for $m_y>0$ is the normalization factor with $m_y$ being the eigenvalue of $\sum_{\mu-\nu=y}\hat{m}_{\mu,\nu}$, namely the total number of $y$-excitons on the chain, which is counted after creation action and before annihilation, and $\zeta_y=0$ for $m_y=0$ as the usual definition. The phase $\Theta$ is determined by external field and without field it is randomly chosen for each $y$. This definition is exact when $L/2$ is divisible by $y$ and can be safely extended to the thermodynamic limit.

Considering the commutation relation (\ref{commu}), we can define normalized Fock states of $y$-excitons, i.e.
\begin{eqnarray}
|\cdot\cdot\cdot0_{y'}\cdot\cdot\cdot m_y\cdot\cdot\cdot0_{y''}\cdot\cdot\cdot)=\frac{1}{\sqrt{m_y!}}\hat{b}^{\dag m_y}_{y}|\circ\rangle,
\end{eqnarray}
where the number of other $y'$-excitons $m_{y'}$ equals to zero, reflecting they are individual substances. For simplicity, in the following we use abbreviation $|m_y)$ while without ambiguity. Then we have
\begin{eqnarray}
&&\hat{b}^{\dag}_{y}|m_y)=\sqrt{m_y+1}|m_y+1),\nonumber\\
&&\hat{b}_y|m_y)=\sqrt{m_y}|m_y-1),\nonumber\\
&&\hat{m}_y|m_y)=\hat{b}^{\dag}_y\hat{b}_y|m_y)=m_y|m_y),
\end{eqnarray}
following the usual form of bosons, and other $\hat{b}_{y'}|0_{y'}\cdot\cdot\cdot m_y)=0$ by definition.

Here, to distinguish from the normal second-quantized representation $|\cdot\rangle$, we use $|\cdot)$ to denote the newly-constructed representation for $y$-excitons. It is actually over complete because its Hilbert space is much larger than $3^L$. For example, the state $|++--\rangle$ can be assigned with two groups $|\uline{+}\uwave{+}\uline{-}\uwave{-}\rangle$ and $|\uline{+}\uwave{+}\uwave{-}\uline{-}\rangle$ equivalently. So it can be represented by either $|0_12_20_3)$ or $|1_10_21_3)$, which exactly have the same particle number and, by our construction below, the same energy. This redundant degree of freedom serves as the essential property of this representation, leading to exotic physics as addressed soon. It also provides rich structure for fragmentation of Hilbert space \cite{frag}, which is valuable for further exploration.

By the Fock states, the coherent state of $y$-excitons in thermodynamic limit can be straightforwardly defined as
\begin{eqnarray}
\hat{b}_y|\lambda_y)=\lambda_y|\lambda_y),~~|\lambda_y)=e^{\lambda_y\hat{b}^{\dag}_y-\frac{1}{2}|\lambda_y|^2}|\circ\rangle,
\end{eqnarray}
where $(\lambda_{y'}|\lambda_y)=\delta_{y,y'}$ and $\hat{b}_{y'}|\lambda_y)=0$ for $y\neq y'$.

In solid materials, a photogenerated exciton is present when there is a finite band gap to absorb the external photon, and there is not an efficient absorption in metallic conductors. In our formalism, $|\lambda_y)$ is not the eigenstate of $\hat{b}^{\dag}_{y}$ as the coherent states are over complete, so the action of creation operator on the coherent state can be regarded as the absorption. As $(\lambda_y|\hat{b}^{\dag}_y|\lambda_y)=\lambda^{*}_y$, meaning that $\lambda^{*-1}_y\hat{b}^{\dag}_y|\lambda_y)$ has large overlap with $|\lambda_y)$, we can think the coherent state is stable against fluctuations and the small perturbation will be quickly relaxed, which is an irreversible process. The excited number of $y$-excitons is thus calculated as
\begin{eqnarray}
\langle\hat{m}_{y}\rangle=\frac{(\lambda_y|\hat{b}_{y}\hat{b}^{\dag}_{y}\hat{b}_{y}\hat{b}^{\dag}_{y}|\lambda_y)}
{(\lambda_y|\hat{b}_{y}\hat{b}^{\dag}_{y}|\lambda_y)}=\frac{1+3|\lambda_y|^2+|\lambda_y|^4}{1+|\lambda_y|^2}.
\end{eqnarray}
Since the average number of the coherent state is $(\lambda_y|\hat{b}^{\dag}_{y}\hat{b}_{y}|\lambda_y)=|\lambda_y|^2$, the absorbed number then equals to $1+|\lambda_y|^2-\frac{|\lambda_y|^4}{1+|\lambda_y|^2}$. The first $1$ refers to the direct photonic gap. The last two terms arise due to the coherence among various Fock states, which can be regarded as the thermal fluctuation along with the absorption, since the fluctuation of coherent state is calculated as $\langle\hat{m}_y^2\rangle-\langle \hat{m}_y\rangle^2=|\lambda_y|^2$.

\subsection{Microcanonical ensemble of $y$-excitons}

A natural question now arises, what is the energy of $y$-exciton? This surely depends on how we write down the Hamiltonian, but it is also a matter of how we choose to express the thermodynamics. At first glance, one might be thinking $y$ as the energy of $y$-exciton, so that photons with frequency $y$ can resonantly excite the $y$-excitons. This is not the case. As stated, if the energy depends on $y$, two identical states $|0_12_20_3)$ and $|1_10_21_3)$ may have different energy. Actually since $y$ is the distance between electron and hole, in practice, we can continuously decrease it via for example decreasing the intermolecular spacing, such that the excitation energy becomes to the order of thermal fluctuation, and any external photons with small frequency can be absorbed, violating the basic rule of photoelectric effect.

Let us explain this in a thermodynamic manner. When $y\leq L/2$, the total number of states, namely the maximum $m_y$, has the order of $L/2$ and even equals to $L/2$ when it is divisible by $y$. That is to say, if we use the Gibbs entropy for each $y$-exciton \cite{negative}, following the usual definition with $k_{\rm B}=1$, it is calculated to be
\begin{eqnarray}
S_{\rm G}\simeq\ln \frac{L}{2}.
\end{eqnarray}
This entropy does almost not depend on $y$ in a large extent, inconsistent with the normal definition. Therefore, $y$ can not be explicitly related to the energy. This is similar to the band theory, namely no matter how large $y$ is, all coherent sites together form an energy band so that the excitons are collective excitations between valence and conduction bands.

Subsequently, we address an essential hypothesis in this paper that, \textit{each $y$-exciton exactly possesses the same excited energy}, which equals to the chemical potential difference between vacuum and electron plus hole. How to account for the issue of semi-classical Coulomb attraction between electron and hole will be discussed below. It is worth noting that, the reason we do not use $y$ as the energy and construct a manifold of exciton above the ground state is benefitting from the advantage of balanced ternary formalism; one can not construct the same manifold in the framework of binary formalism.

For simplicity, in the following we set the chemical potential difference between vacuum and excited states to be unity, so the total number of particle $m$ is equivalent to the total energy. They also share the same fluctuation of the order $|\lambda_y|^2$ in the excited manifold. The total number of possible states becomes around $m^{L/2}$ for $y\leq L/2$, and we calculate the Boltzmann entropy as
\begin{eqnarray}
S_{\rm B}\simeq\frac{L}{2}\ln m,
\end{eqnarray}
constituting a convex function of energy. Then the Boltzmann temperature for microcanonical ensemble of each energy surface turns out to be
\begin{eqnarray}
T_{\rm B}=(\frac{\partial S_{\rm B}}{\partial m})^{-1}\simeq\frac{2m}{L},
\end{eqnarray}
which is the density of excitons.

Notice that, when $y>L/2$ the possible number of that exciton is decreased. For instance, $y=L-1$ corresponds to only one state $|+\cdot\cdot\cdot-\rangle$. The largest particle number is then negatively proportional to $y$ for $y>L/2$, i.e. $m\sim(L-y)$, but the average energy is still the same for all $y$'s. Hence, the consideration of $y>L/2$ cases will still increase the entropy.

\subsection{Canonical ensemble of $y$-excitons}

We turn to consider the canonical ensemble of chains with length $L$. It is assumed that there are many thermal photons with mean energy $m$ that may thermally excite $y$-excitons. Once an exciton is created on some sites, those sites can no longer be occupied by other excitons, namely $\hat{b}^{\dag}_{y'}|0_{y'},\lambda_y)=|1_{y'},\lambda_y)$ is solely not orthogonal to $|\lambda_{y'},\lambda_y)$, and only the relevant annihilation operator can eliminate it, except the high-order processes as discussed below. So if the chain is acted by an external photonic environment with repeated creations and annihilations, the chain will be randomly occupied by mixture of various excitons, until the detailed balance is reached. In addition, in our theory we can even allow the superposition of various $y$-excitons, one photon $m=1$ can also simultaneously excite several $y$-excitons with fractal concentrations. The question then arises: how can we determine the final thermal state of the ensemble? Or what state is the stablest to which any excited states induced by perturbation of external environment would rapidly relax?

A first glance may give rise to the normal thermal state as $\exp(-\beta \hat{H})$, with $\hat{H}\sim\hat{b}^{\dag}_y\hat{b}_y$. In our viewpoint, however, this form is not available for excitons, since different from other bosons, excitons are strongly interacting with environment leading to large fluctuations of particle number and energy. In the strong-coupling limit, therefore, we neglect the frequency term and turn to determine the eigenstate of stabilizer
\begin{eqnarray}
\hat{S}=\sum_y(\hat{b}_y+\hat{b}^{\dag}_y).
\end{eqnarray}
That is to say, the thermal photons and phonons are all assumed to be linearly coupled with excitons on the polarization direction. The differences of coupling strengths are neglected due to the hypothesis that the energies of various excitons are all the same. This assumption follows the most common sense of light-matter interactions and, in our opinion, can be applied to most excitonic systems. We further notice that, $[\hat{b}_y+\hat{b}^{\dag}_{y'},\hat{b}_{y'}+\hat{b}^{\dag}_{y}]=0$ for any case, so we can regroup the creation and annihilation operators and divide the stabilizer into two terms
\begin{eqnarray}
\hat{S}=\hat{B}+\hat{B}^{\dag},~~\hat{B}=\sum_{\rm odd~\it y}(\hat{b}_y+\hat{b}^{\dag}_{y+1}).
\end{eqnarray}
As $[\hat{B},\hat{B}^{\dag}]=0$, they share the same eigenstates, which can then be the candidate of the stabilized state and the thermal state.

In order to construct these eigenstates, one advantage of our balanced ternary form then turns out. The vacuum state has an inverse case, namely the fully occupied state $|\bullet\rangle$ that the chain is occupied by excitons as most as possible. This can be realized either in gapless metals, that near the Fermi level almost all the states of electron and hole should be occupied, even if the temperature is very low, or in the exciton condensate \cite{conden1,conden2}. Note that, here the words ``fully occupied" are meant to excitons, not to electrons and holes. Then we can construct a complementary form of coherent state as the eigenstate of creation operator,
\begin{eqnarray}
\hat{b}^{\dag}_y|\tilde{\lambda}_y)=-\tilde{\lambda}_y|\tilde{\lambda}_y),~~|\tilde{\lambda}_y)=e^{\tilde{\lambda}_y\hat{b}_y+\frac{1}{2}|\tilde{\lambda}_y|^2}|\bullet\rangle.
\end{eqnarray}
Obviously,
\begin{eqnarray}
(\tilde{\lambda}_y|\hat{b}_y\hat{b}^{\dag}_y|\tilde{\lambda}_y)=|\tilde{\lambda}_y|^2.
\end{eqnarray}
Different from the canonical coherent state, here in the complementary coherent state the larger the $\tilde{\lambda}_y$, the smaller the particle number and the energy. In order to make $\langle\hat{b}^{\dag}_y\hat{b}_y\rangle>0$, $|\tilde{\lambda}_y|^2$ should be always larger than one, so the distribution is still concentrated to the low-energy states in normal semiconductors. When $|\tilde{\lambda}_y|^2-1=|\lambda_y|^2$, these two coherent states have the same mean energy. As the eigenstate of creation operator, it can not absorb additional energy from external photons.

Given these two types of coherent states, we can now use for example the product coherent state as the eigenstate of the stabilizer, i.e.
\begin{eqnarray}\label{thermal}
|T_{\rm E}\rangle=\prod_{\rm odd~\it y}|\lambda_y\tilde{\lambda}_{y+1})=|\cdot\cdot\cdot\lambda_{y}\tilde{\lambda}_{y+1}\cdot\cdot\cdot),
\end{eqnarray}
which is expected to be invariant under the action of repeated creation and annihilation. If we do not consider the specific crystalline and topological structures, the index $y$ can be arbitrarily reordered to generate a degeneracy $(L-1)!/(((L-1)/2)!)^2$. Each degenerate configuration should take the same probability to fulfill the equipartition theorem. The entropy of the thermal state then equals to $(L-1)\ln2$, similar to the total entropy of the original fermion system, meaning that no additional information has been generated by regrouping the excitons, as expected. While considering spin degree of freedom, as discussed soon, the degeneracy will be decreased.

It is worth noting that, $\hat{b}_y$ and $\hat{b}^{\dag}_{y'}$ are generally not commutative while acting on any states other than vacuum, so the reordering of coherent states in (\ref{thermal}) may give rise to an additional phase. However, considering the whole chain is occupied by various excitons, the mean particle number for each $y$ should be smaller than one. Further considering an approximation of the random phase $\Theta$ for each $y$, the noncommutation of $\hat{b}_y$ and $\hat{b}^{\dag}_{y'}$ and their ordering will not significantly influence the practical calculations.

In order to determine the canonical temperature $T_{\rm E}$, the set of $\lambda_y$ and $\tilde{\lambda}_y$ is essential. Considering they are continuous variables, different set of them has great chance to give different summation. Hence we can solely discuss the distribution of the total energy or particle number, instead of individual $\lambda_y$. An intuitive idea is to make the excitons be equilibrium with an external black-body radiations, so that the total particle number satisfies Bose-Einstein distribution. We have
\begin{eqnarray}\label{BE}
&&m=\sum_y\langle \hat{m}_y\rangle=\sum_{\rm odd~\it y}(|\lambda_y|^2+|\tilde{\lambda}_{y+1}|^2-1)\nonumber\\
&&=\frac{1}{e^{\beta_{\rm E}}-1},
\end{eqnarray}
where $\beta_{\rm E}=1/T_{\rm E}$ is the inverse temperature of external environment. By this definition, $\lambda_y$ and $\tilde{\lambda}_y$ can be set with respect to the canonical temperature. In high-temperature limit, $\lambda_y$, equivalent to the displacement of exciton, is proportion to $T_{\rm E}^{1/2}$, also fulfilling the equipartition theorem. As it increases following temperature increasing, the excitons could play a role of thermal medium for charge carriers.

\section{Many-body interactions}

The advantage of the present balanced ternary formalism is that, there are three substances in the framework: Electron, hole and $y$-exciton. There are several sets of noncommutative operators of them. The many-body interactions can be more explicitly expressed than that in binary formalism. In this Section, we briefly discuss some typical instances.

\subsection{Coulomb interaction}

In our theory, the thermal state (\ref{thermal}) is generic, and $\lambda_y$ and $\tilde{\lambda}_{y}$ are the main variational parameters. In metals and exciton condensates, the concentration of excitons is large, so $\lambda_y$ is large and $\tilde{\lambda}_{y}$ is small. Various $y$-excitons should be mixed with each other as even as possible to produce the largest entropy, and there is not a specific distribution for $y$. This can be understood as the Coulomb interaction between electron and hole is greatly screened in this large concentration systems. In normal semiconductors without exciton condensation, on the other hand, the concentration is low so that $\lambda_y$ is small and $\tilde{\lambda}_{y}$ is large. This is also understandable as the dielectric constant in semiconductors is large such that the capture radius (the radius that the Coulomb interaction is larger than the thermal energy) should be much shorter than the coherent length of excitons.

In spite of this, there is still a spontaneous tendency that the exciton prefers to change its spatial extent. Let us consider the phase $\Theta$'s are all close to vanish, i.e. there is an electric field to polarize all excitons. When two excitons have overlap in space, we have
\begin{eqnarray}\label{reset}
\hat{a}^{\dag}_{\mu,\nu}\hat{a}^{\dag}_{\mu',\nu'}=-\hat{a}^{\dag}_{\mu,\nu'}\hat{a}^{\dag}_{\mu',\nu}.
\end{eqnarray}
Namely, for example, $|\uline{+}\uwave{+}\uline{-}\uwave{-}\rangle$ and $|\uline{+}\uwave{+}\uwave{-}\uline{-}\rangle$ are equivalent. According to the entropy increase principle, two same excitons tend to be regrouped to two different ones to occupy more states, and one of them must be with small $y$ and the other with large $y$. As a result, more and more different excitons are generated until reaching the shortest one $y=1$ or the longest one $y=L-1$. This argument can be even stronger in high dimensions. Subsequently, the shrink branch can be regarded as the Coulomb attraction and the spread branch refers to the dissociation of excitons. The shortest excitons can not be further regrouped in this way, so they will have more concentrations than others indicating the equivalent effect of Coulomb attraction. On the other hand, the longest exciton merely has a maximum number one, so when more electrons and holes concentrate on the ends of the chain, they can only incoherently recombine with charge carriers on other chains leading to the charge separation in solar cells. Further extraction of the charge carriers will be discussed in the next Subsection.

Except for the electron-hole attraction, the electron-electron repulsion which can be mapped to the spin exchange energy via Jordan-Wigner transformation, can also be understood in this perspective. The repulsive interaction in spinless fermion chain normally induces the charge-density wave (CDW), like $|+-+-+-\rangle$, which has also the relatively high concentration in our setup. This configuration can be realized as the exciton condensation in CDW as well \cite{conden1,conden2}.

\subsection{Exciton-facilitated charge current}

We further notice that, $\hat{f}^{\dag}_r\hat{a}^{\dag}_{\mu,\nu}|\circ\rangle$ and $\hat{g}_{r'}\hat{a}^{\dag}_{\mu,\nu}|\circ\rangle$ are merely nonvanishing when $r=\nu$ and $r'=\mu$. That is,
\begin{eqnarray}
&&\hat{f}^{\dag}_{\nu}\hat{g}^{\dag}_{\mu}\hat{f}^{\dag}_{\mu}\hat{f}_{\mu}\hat{\xi}\hat{f}_{\nu}\hat{g}_{\nu}\hat{g}^{\dag}_{\nu}|\circ\rangle=-\hat{g}^{\dag}_{\mu}|\circ\rangle,\nonumber\\
&&\hat{g}_{\mu}\hat{g}^{\dag}_{\mu}\hat{f}^{\dag}_{\mu}\hat{f}_{\mu}\hat{\xi}\hat{f}_{\nu}\hat{g}_{\nu}\hat{g}^{\dag}_{\nu}|\circ\rangle=\hat{f}_{\nu}|\circ\rangle.
\end{eqnarray}
Then it is not difficult to derive that,
\begin{eqnarray}\label{fb}
&&\hat{f}^{\dag}_{r}|\lambda_y)=-\lambda_y\hat{g}^{\dag}_{r+y}|\lambda_y),\nonumber\\
&&\hat{g}_{r'}|\lambda_y)=\lambda_y\hat{f}_{r'-y}|\lambda_y).
\end{eqnarray}
This result is essential to comprehend the hopping of electron and hole in a thermal environment. Namely, the electron or hole is injected onto some site of the chain, the coherent state of $y$-excitons will play the role of medium to produce a dispersive wave of image electrons or holes away from the origin, similar to the mechanism of hopping. The larger the $\lambda$ and the higher the temperature $T_{\rm E}$, the stronger the hopping.

Rather, $\hat{f}^{\dag}$ and $\hat{g}$ do not represent a real electron and hole; we have to consider the hopping of real particles. We further notice that, the coherent state of $y$-excitons has not got net electron or hole, so the mean chemical potential $\alpha$ equals to zero. The Bogoliubov transformation $\exp(i\theta\hat{\gamma})$ can however act on $|\pm\rangle$ states to generate net electron or hole. It is then derived that
\begin{eqnarray}
e^{i\theta_r\hat{\gamma}_r}|\lambda_y)=|\lambda_y;+_{r+y},\theta_{r}\rangle+|\lambda_y;\theta'_{r},-_{r-y}\rangle,
\end{eqnarray}
where all sites are at coherent state except there is an unpaired electron at site $r+y$ and a hole at $r-y$, and $|\theta\rangle=\cos\theta|-\rangle+\lambda_y\sin\theta|+\rangle$, $|\theta'\rangle=-\lambda_y\sin\theta|-\rangle+\cos\theta|+\rangle$.  When $\theta=\pi/2$, we can obtain pure $|++\rangle$ and $|--\rangle$ states in the chain, which can be named as negative and positive diradicals, respectively. So the generation of net electron or hole depends on the chemical potential with the relation $\exp(-\alpha)$. Most importantly, we can not construct a set of operators to directly generate diradicals from vacuum. The coherent states of excitons are essential, as without their nonconservation of particle numbers there will yet be any net charges. This mechanism can also be thought as that the electrons and positrons are very initially generated from photons.

More interestingly, as the operator $\hat{b}^{\dag}_{y}$ with $\Theta=0$ can not generate $|-\rangle$ on the largest $y$ sites, meaning in a thermal equilibrium these sites can only be occupied by $|+\rangle$, so we can call them electron-rich region. Equivalently, the smallest $y$ sites can only be occupied by $|-\rangle$, so they are hole-rich region. This charge polarization may produce a built-in field further enhancing the charge accumulation, which is the normal case in semiconducting devices or the photovoltage in solar cells.

A net charge current can then be generated by recombining these rich electrons or holes. This can be realized by applying $\exp (i\theta\hat{\gamma}_r)$, since in the region $r>L-y$ merely $|--\rangle$ state can be induced, and equivalent case is for the region $r\leq y$. Essentially, the actions of $\exp (i\theta\hat{\gamma}_r)$ in these regions are irreversible, giving rise to a steady flow of charge current. With the action of $\exp (i\theta\hat{\gamma}_r)$ on the hole-rich region, if we set $d\theta/dt=\omega$ constant, the temporal derivative of total number of particle then writes
\begin{eqnarray}
\frac{dN}{dt}=\frac{\omega\sin 2\theta}{1-e^{-\beta_{\rm E}}},
\end{eqnarray}
which can be regarded as the ac current, and there is not a net dc current when $\omega=0$.

If we want to generate a dc current, it is necessary to concurrently consider another action on the electron-rich region, i.e.,
\begin{eqnarray}
&&e^{i(\theta_L\hat{\gamma}_L+\theta_1\hat{\gamma}_1)}|\lambda_y)\simeq-\lambda_y\sin\theta_{L}\cos\theta_{1}|-_{L},\lambda_y,-_{1}\rangle\nonumber\\
&&-\lambda_y^2\sin\theta_L\sin\theta_1|-_{L},\lambda_y,+_{1}\rangle\nonumber\\
&&+\cos\theta_L\cos\theta_1|+_{L},\lambda_y,-_{1}\rangle\nonumber\\
&&+\lambda_y\cos\theta_{L}\sin\theta_{1}|+_{L},\lambda_y,+_{1}\rangle.
\end{eqnarray}
Here we use $\simeq$ because in the middle region there should be some unpaired electron and hole, not exactly the coherent state, but this does not matter in the following calculation. So we redefine the current operator (\ref{current}) as
\begin{eqnarray}
\hat{J}_{L,1}=-i(e^{i\phi}\hat{f}_{L}\hat{g}_{L}\hat{g}^{\dag}_{1}\hat{f}^{\dag}_{1}-e^{-i\phi}\hat{f}_{1}\hat{g}_{1}\hat{g}^{\dag}_{L}\hat{f}^{\dag}_{L}).
\end{eqnarray}
In terms of gauge theory, an additional phase $\phi$ is introduced to reflect the effect of an external electric field, which is positive for electrons in the hole-rich region and holes in the electron-rich region and for the reverse case it is negative. Then the dc current can be calculated as
\begin{eqnarray}
&&\langle\hat{J}_{L,1}\rangle=2\lambda^2_y\sin\phi\sin\theta_L\cos\theta_L\sin\theta_1\cos\theta_1\nonumber\\
&&=\frac{\lambda^2_y\sin\phi}{2\cosh\alpha_L\cosh\alpha_1}.
\end{eqnarray}
From this formula, the current is generated by two steps. The first is to produce electron and hole by $\hat{\gamma}$ on the ends of the chain. The second is to add a phase $\phi$ to move them onto another end.

The result is essential in our present theory, as it uniformly gives two types of temperature dependency. In the low-temperature limit, $\lambda_y$ is solely nonzero in the near neighborhood and the excitons are localized, so the temperature dependency is dominated by $\alpha=-\mu/k_{\rm B}T$. Integrating from 0 to $\mu$, the current will be roughly proportional to $T^2$, so following $T$ increasing the current decreases, giving rise to a negative temperature coefficient. In organic semiconductors this is the bandlike transport \cite{band1,band2}. On the other hand, in the high-temperature limit, $\alpha\simeq0$ and $\langle\hat{J}_{L,1}\rangle=\frac{1}{2}\lambda^2_y\sin\phi$, so the mean distance in each hopping follows $\lambda^2_y$ to increase with temperature increasing, resulting in positive temperature coefficient, which is the hopping transport. The bandlike-hopping transition long-termly serves as the puzzle in organic electronics \cite{CT}, and here we give a possible explanation. More detailed analysis of distribution on $y$ is remained for numerical computations.

\subsection{Exciton fission}

There are many interaction processes of excitons, such as singlet fission and triplet-triplet annihilation \cite{SF}. The first is to split one singlet exciton into two entangled triplet excitons. In literature, people commonly thought this process by considering the energy of singlet is twice of the triplet energy. The most exotic effect, namely the nonconservation of exciton particle number, has not been comprehensively addressed. Just like that we can not directly split one photon into two, it is of course significant to figure out an intrinsic mechanism for this nonconservation.

Although we can reset the groups as Eq.~(\ref{reset}), the total number of particles is the same. The change of particle number normally relies on the large number of particles, i.e. the collective effect in condensed matter. When a creation operator is acted on its eigenstate, i.e. the complementary coherent state, nothing changes, but if it acts on the canonical coherent state, it will try to transfer the energy to other complementary states. Interestingly, let us consider a special case that the coherent length is very short, for example the maximum $y$ is two. Then there are only two thermal states $|\tilde{\lambda}_1,\lambda_2)$ and $|\lambda_1,\tilde{\lambda}_2)$. Now if a $y=3$ exciton is generated, it is not stable and will be relaxed to the thermal state. The sole action turns out to be
\begin{eqnarray}\label{fission}
&&\hat{a}^{\dag}_{r,r+3}|\tilde{\lambda}_1,\lambda_2)=\hat{a}^{\dag}_{r,r+3}(\frac{-\hat{b}^{\dag}_1}{\tilde{\lambda}_1})|\tilde{\lambda}_1,\lambda_2)\nonumber\\
&&\rightarrow\frac{\zeta}{\tilde{\lambda}_1}\hat{a}^{\dag}_{r,r+2}\hat{a}^{\dag}_{r+1,r+3}|\tilde{\lambda}_1,\lambda_2).
\end{eqnarray}
The arrow on the second line indicates the process is not exactly energy conserved. Considering the energy of $y=1$ does not change, the energy change during this process is roughly $\frac{2}{|\tilde{\lambda}_1|^2}-1$. For $|\tilde{\lambda}_1|>\sqrt{2}$ it is an exothermal process, while for $1<|\tilde{\lambda}_1|<\sqrt{2}$ it is endothermal. Most importantly, there appear two emergent $y=2$ excitons that can not be absorbed into the thermal state, as $|\lambda_2)$ is canonical coherent state. So we can design some mechanism to extract two electrons and holes. Hence, we can call this process as exciton fission. It takes place in a material of trimer and will be weakened in tetramer \cite{trimer}. The reversed process, namely the fusion of two excitons, can also be discussed in this scenario.

\section{Qutrits with spin half}

In above discussions, we neglect the spins of electrons/holes and also excitons. Electrons and holes possess spin half, and excitons can be either singlet or triplet. But in most inorganic materials, the spin-orbital coupling is sufficiently strong to hybridize the spin states, so the above discussions are available in these situations. In organic semiconductors, however, the light elements can not efficiently change the spin states. The direct photon excitation generates a singlet exciton which can merely be transformed to triplet via intersystem crossing and other similar mechanisms. The chemical potential of triplet can be totally different with that of singlet. In singlet fission as discussed above, a triplet has an energy half of singlet energy. So we have to conduct a mechanism of transition between singlet and triplet to enforce the exciton fission process.

Of course one can analogously construct a ternary formalism, like the present one, for the triplet excitons. But that means the charge and spin degrees of freedom are completely decoupled. Here alternatively, we will universally consider both charge and spin in a uniform framework. To this end, we can simply rearrange the site index as
\begin{equation}
\cdot\cdot\cdot,2\uparrow,2\downarrow,1\uparrow,1\downarrow,
\end{equation}
where the number denotes the spatial sites and $\uparrow$ and $\downarrow$ for spin up and down, respectively. Here, the spin up of hole means there is an unpaired electron with spin down remaining on its orbital, which is able to recombine with the electron with spin up, so the electron and hole with the same spin state are still exclusive in the same site. Other arrangement of site and spin index may lead to confused exciton index, so this arrangement is unique.

In this setup, neglecting the on-site excitons, there emerge four modes of excitons, $|\uparrow\uparrow\rangle$, $|\downarrow\downarrow\rangle$, $|\uparrow\downarrow\rangle$ and $|\downarrow\uparrow\rangle$. If we construct a ladder with the upper leg for spin up and lower leg for spin down, these four modes can be labeled as two parallel modes $|\rightharpoonup)$, $|\rightharpoondown)$, and two crossed modes $|\searrow)$, $|\nearrow)$, respectively. The arrows point from electron to hole. The $y$-exciton is then renamed as $y,s$-exciton denoted by $|y,s)$, with $s$ taking these four spin modes. Since the spin state of hole is reversed with that of electron, two parallel modes can be regarded as the mixture of singlet S and triplet T$_0$, while the crossed two are triplets T$_+$ and T$_-$. So the above exciton fission process (\ref{fission}) should be thought as
\begin{eqnarray}
|2\rightharpoonup)-|1\rightharpoondown)\rightarrow|1\searrow)+|2\nearrow),
\end{eqnarray}
equivalent to the singlet fission.

The energy of excitons with various modes can now be different, since the groups $|\uline{+}\uwave{+}\uline{-}\uwave{-}\rangle$ and $|\uline{+}\uwave{+}\uwave{-}\uline{-}\rangle$ are no longer equivalent. We then have to assign different excitation energy (band gap) and transition dipole moment for $s$-excitons individually. In the thermal state (\ref{thermal}) we have already distinguished $y$ by using canonical and complementary coherent states, which can be straightforwardly applied to the parallel and crossed spin modes. Normally in organic semiconductors, the concentration of triplet should be much smaller than that of singlet, and we can use $\lambda_{y,s}$ and $\tilde{\lambda}_{y,s}$ to control the concentrations.

\subsection{Ferromagnetic ordering}

\begin{figure}[t]
    \centering
    \includegraphics[width=0.9\linewidth]{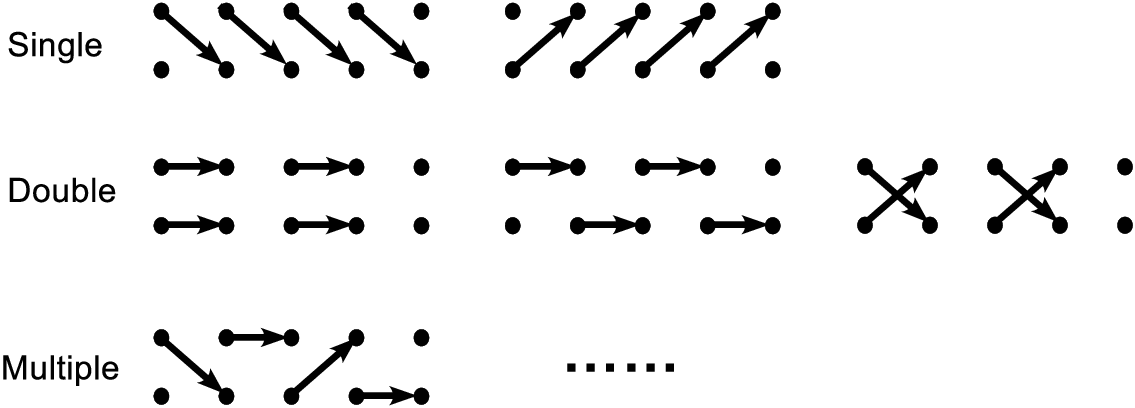}
    \caption{\label{config1} Possible configurations of nearest-neighbour excitons with single, double and multiple modes.}
\end{figure}

We now study the fully occupied state $|\bullet\rangle$ of the shortest $y$-excitons, i.e. the electron and hole reside on the nearest-neighbor sites. The other cases can be similarly discussed. The possible inequivalent configurations of spin modes are summarized in Fig.~\ref{config1}. We can classify these configurations into three types: Single, double and multiple modes. There are solely two configurations for the single-mode type, that is, all sites are occupied by one of the crossed modes individually. Each parallel mode can not fill the whole chain by itself, so two configurations consisting of double modes are mixed with two parallel modes. There are also a configuration of crossed two modes and more diverse ones of multiple modes which are not completely listed. By the ergodic hypothesis, all these configurations should share the same probability. But different from the above spinless situations in which all excitons with the same $y$ are identical, here the configurations possess completely different spin modes, and those with large gap should be stabler than gapless ones.

The most appealing result then points to the single-mode configurations, which can be regarded as two ferromagnetic orderings, as the spin down of hole is equivalent to spin up of electron. One can find that, in order to change the local spin modes of the configuration, it is necessary to transform two nearest excitons, namely transform $|\searrow\searrow)$ or $|\nearrow\nearrow)$ to two parallel modes, respectively. Other configurations are all spin gapless, since this transformation does not change the mean value of spin. That is to say, only the two configurations of single modes have got a finite spin gap, which protects the ferromagnetic orderings at finite temperature. The spin gap can be estimated by annihilating two excitons from the complementary coherent state, namely
\begin{eqnarray}
&&\langle\hat{b}^{\dag}_{1,s}\hat{b}_{1,s}\rangle=\frac{(\tilde{\lambda}_{1,s}|\hat{b}^{\dag}_{1,s}\hat{b}^{\dag}_{1,s}\hat{b}^{\dag}_{1,s}\hat{b}_{1,s}\hat{b}_{1,s}\hat{b}_{1,s}|\tilde{\lambda}_{1,s})}
{(\tilde{\lambda}_{1,s}|\hat{b}^{\dag}_{1,s}\hat{b}^{\dag}_{1,s}\hat{b}_{1,s}\hat{b}_{1,s}|\tilde{\lambda}_{1,s})}\nonumber\\
&&=\frac{-6+18|\tilde{\lambda}_{1,s}|^2-9|\tilde{\lambda}_{1,s}|^4+|\tilde{\lambda}_{1,s}|^6}{2-4|\tilde{\lambda}_{1,s}|^2+|\tilde{\lambda}_{1,s}|^4}.
\end{eqnarray}
The average energy becomes positive definite when $|\tilde{\lambda}|>2.5$, meaning a finite spin gap robustly exists at low temperature as the larger the $|\tilde{\lambda}|$ the lower the temperature. So we can use $|\tilde{\lambda}_{\rm c}|=2.5$ to estimate the Curie temperature of the ferromagnetic orderings in the realistic materials.

Different from the normal ferromagnetism stemming from direct spin exchange or mediated by conducting electrons, the present ferromagnetic ordering is induced by the specific configurations of CT excitons. In practice, this relies on a sufficiently large concentration of electrons and holes, which can be obtained by heavily doping. When we synthesized the organic ferromagnetic materials \cite{AM}, e.g., we fould that the optimal way is firstly reducing the neutral molecule to divalent, and then oxidizing it back to monovalent and neutral states. By this treatment, there appears a mixture of three valences and the large concentration is possibly realized, so is the room-temperature ferromagnetism.

\subsection{Magnetoresistance}

As stated, in the present formalism, the charge current is primarily a boundary effect in the excitonic chain. In Fig.~\ref{config1} we can find the single-mode configurations always have two empty sites on both ends of the chain which can be occupied by additional charge carriers. Under an external magnetic field, the probability of these ferromagnetic ordering should be increased so that more empty sites on the ends are induced leading to increase of charge injection. This is the negative magnetoresistance. On the other hand, with the increase of external field, the longest exciton $y=L-1$ can not be flipped by $\exp(i\theta\hat{\gamma})$, due to the presence of spin gap, so the charge current will be decreased. This is the positive magnetoresistance. Both mechanisms essentially rely on the entanglement dynamics, in agreement with our previous work \cite{siwei}. The concrete investigation requiring external field is out of the scope of present work, so it is remained for future researches.

\section{Is there a Hamiltonian?}

What is the typical Hamiltonian in the balanced ternary formalism? Can we write down the usual hopping term of electrons like $\hat{c}^{\dag}_{\mu}\hat{c}_{\mu+1}+{\rm h.c.}$? Surely not. Due to the presence of recombination between electron and hole, we exclude the situation they reside on the same site with the same spin. As there should be a lot of electrons and holes randomly distributing on the chain, it is quite easy for them to meet while hopping. So we can not think the Hamiltonian has the hopping term to produce a plane wave as the eigenstate of electrons or holes. Actually, based on the balanced ternary formalism, we can only realize the natural shape of them is localization on a single site. This also fulfills the rational perspective of charge carriers in a thermal environment.

Can we alternatively write down the usual Hamiltonian of harmonic oscillator like $\omega\hat{b}^{\dag}_{y}\hat{b}_{y}$ for $y$-excitons? Also not. The exciton is a hard-core boson exclusively with each other, so we can not well define its coordinate and momentum. It is commonly generated and held by external photonic and thermal environment, so its creation and annihilation, linear with the transition dipole, are much more important than the occupation. This is why we think it should be empirically stabilized by $\hat{b}^{\dag}_{y}+\hat{b}_{y}$.

In condensed matter theory, especially in superconductors, the particle number is conjugated to the phase, so they satisfy the uncertainty relation. One may thus write down a term with respect to the pairing mechanism between electron and hole. There are also some models for high-$T_{\rm c}$ superconductivity that are related to interactions between pairing bonds. The typical instance is the quantum dimer model \cite{dimer}, which can be equivalent to the model of resonating valence bond (RVB). Another instance is the AKLT Hamiltonian \cite{AKLT}, which involves projectors of spin 2 formed by four spin halves. Different from all these Hamiltonians, our theory mainly takes the coherent length of CT excitons into account, so it is difficult to straightforwardly borrow their forms.

Actually, our present scope is mainly focusing on the quantum thermodynamics in the fermionic chain. Although the formalism is compact, it is not conservative, and there are not explicitly conserved quantities such as the energy and particle number. This is because we primarily use the coherent states as the thermal states, which have got large fluctuations of energy and particle number. It agrees with the realistic situations as the excitons are continuously created and annihilated to achieve the detailed balance. In this context, people normally use projectors to formalize the Hamiltonian. We can also follow this way, if necessary, but we prefer to enumerate all possible actions rather than writing down a Hamiltonian. This treatment can be more friendly to the quantum control and simulation to form the quantum circuits.

\section{Conclusions}

In summary, we have constructed a second-quantized representation based upon the balanced ternary numbers. In this formalism, we consider three substances: Electron, hole and $y$-excitons. By bosonization on a chain, we construct the creation and annihilation of excitons, as well as their thermodynamics. Via exchanging the operators of different $y$-excitons, the various interactions among these substances are investigated. The charge current and exciton fission are quantified by the coherent states of excitons. The spin half is further considered, and we find two exotic spin configurations that have got spin gap for the presence of ferromagnetic orderings.

The present formalism is applicable mainly in the rigid-backbone disordered materials that the coherent length is much longer than localization length. The crystalline structure, topology and valley polarization must play significant role and will be investigated in the future. Except for the instances discussed in this paper, much more states of matter in electric, magnetic and optical areas are remained for further researches. In particular, we are able to study the energy band and localized orbital in a universal framework. All in all, the balanced ternary formalism has a great extensibility in crossing fields.

\begin{acknowledgments}

The authors gratefully acknowledge support from the National Natural Science Foundation of China (Grant No.~12374107).

\end{acknowledgments}

\newpage

\end{document}